\documentclass[conference]{IEEEtran}
\IEEEoverridecommandlockouts

\usepackage{amsmath}
\usepackage{amssymb}
\usepackage{array}
\usepackage{fixmath}
\usepackage{graphicx}
\usepackage{cite}
\usepackage{color}
\usepackage{changes}
\usepackage{algorithm}
\usepackage{algorithmic}
\usepackage{acronym}
\usepackage{subcaption}
\usepackage{bm}

\usepackage{multirow}

\DeclareMathAlphabet{\mathbit}{OML}{cmr}{bx}{it}

\newacro{TDD}{time-division duplexing}
\newacro{CSI}{channel state information}
\newacro{DL}{downlink}
\newacro{UL}{uplink}
\newacro{BS}{base station}
\newacro{MS}{mobile station}
\newacro{MSE}{mean square error}
\newacro{MMSE}{minimum mean square error}
\newacro{SVD}{singular value decomposition}
\newacro{AM}{alternating minimization}
\newacro{OFDM}{orthogonal frequency-division multiplexing}
\newacro{mmWave}{millimeter wave}
\newacro{OMP}{orthogonal matching pursuit}
\newacro{MIMO}{multiple-input multiple-output}
\newacro{RF}{radio-frequency}
\newacro{LS}{least squares}
\newacro{MRC}{maximum ratio combiner}
\newacro{ZF}{zero-forcing}
\newacro{CS}{compressive sensing}
\newacro{ULA}{uniform linear array}
\newacro{ADC}{analog-to-digital converter}
\newacro{AoA}{angles-of-arrival}
\newacro{AoD}{angles-of-departure}
\newacro{CRLB}{Cram\'{e}r-Rao lower bound} 
\newacro{NMSE}{normalized mean squared error}
\newacro{CB}{coordinated beamforming}
\newacro{NMSE}{normalized mean-squared error}
\newacro{SINR}{signal-to-interference-plus-noise ratio}
\newacro{LLF}{log-likelihood function}
\newacro{SW-OMP}{simultaneous weighted - orthogonal matching pursuit}
\newacro{SS-SW-OMP+Th}{subcarrier-selection simultaneous weighted - orthogonal matching pursuit + thresholding}

\newcommand{\Nr}{N_{\rm R}}
\newcommand{\Nt}{N_{\rm T}}
\newcommand{\nr}{n_{\rm R}}
\newcommand{\nt}{n_{\rm T}}
\newcommand{\Mt}{M_{\rm T}}
\newcommand{\Mr}{M_{\rm R}}
\newcommand{\FRFm}{{\bf F}_{m}^{\rm RF}}
\newcommand{\FBBm}{{\bf F}_{m}^{\rm BB}}
\newcommand{\WRFm}{{\bf W}_{m}^{\rm RF}}
\newcommand{\WBBm}{{\bf W}_{m}^{\rm BB}}
\newcommand{\Fm}{{\bf F}_{m}}
\newcommand{\Wm}{{\bf W}_{m}}
\newcommand{\MrQ}{\Mr Q}
\newcommand{\Ns}{N_{\rm s}}
\newcommand{\at}{{\bf a}_{\rm T}(\boldsymbol \phi_l)}
\newcommand{\atH}{{\bf a}_{\rm T}^{\rm H}(\boldsymbol \phi_l)}
\newcommand{\ar}{{\bf a}_{\rm R}(\boldsymbol \theta_l)}

\def\b0{{\mathbf{0}}}

\def\bN{{\mathbf{N}}}

\def\bS{{\mathbf{S}}}

\def\bY{{\mathbf{Y}}}

\def\sf0{{\mathsf{0}}}

\def\bsf0{{\bm{\mathsf{0}}}}

\begin{document}

\title{Low complexity joint position and channel estimation at millimeter wave based on multidimensional orthogonal matching pursuit}

\author{\IEEEauthorblockN{Joan Palacios,~Nuria Gonz\'alez-Prelcic}\\ 
	\IEEEauthorblockA{%
		\textit{North Carolina State University, USA}\\
		Email:\{\texttt{jbeltra,ngprelcic}\}\texttt{@ncsu.edu}}
\and
\IEEEauthorblockN{Cristian Rusu}
\IEEEauthorblockA{\textit{} \\
\textit{University of Bucharest, Romania}\\
Email:\texttt{cristian.rusu@unibuc.ro}}
}

\maketitle

\begin{abstract}
Compressive approaches provide a means of effective channel high resolution channel estimates in millimeter wave MIMO systems, despite the use of analog and hybrid architectures. Such estimates can also be used as part of a joint channel estimation and localization solution. Achieving good localization performance, though, requires high resolution channel estimates and better methods to exploit those channels. In this paper, we propose a low complexity multidimensional orthogonal matching pursuit strategy for compressive channel estimation based by operating with a product of independent dictionaries for the angular and delay domains, instead of a global large dictionary.  This leads to higher quality channel estimates but with lower complexity than generalizations of conventional solutions. We couple this new algorithm with a novel localization formulation that does not rely on the  absolute time of arrival of the LoS path and exploits the structure of reflections in indoor channels. We show how the new approach is able to operate in realistic 3D scenarios to estimate the communication channel and locate devices in an indoor simulation setting.
\end{abstract}

\section{Introduction}

Millimeter wave (mmWave) communication and MIMO technology offer additional benefits beyond high data rate communications. The large arrays at high frequencies
provide the angle and delay resolvability that enables accurate localization of users and objects in the environment as a byproduct of communication \cite{Shahmansoori2018}. 
Future 5G releases and 6G envision VR/AR and automated cars as main use cases, expecting an indoor accuracy $< 1$ cm and  an outdoor accuracy $< 10$ cm. New research work is needed, from an algorithmic perspective, to achieve this performance.

One direction for developing high accuracy localization in a mmWave MIMO system is to take a model-driven two-stage approach \cite{Shahmansoori2018,Wymeersch2018,Talvitie2019,Jiang2021}. The first stage performs compressive channel estimation, which has proven successful in mmWave and massive MIMO channels explicitly, since the large bandwidth or number of antennas prevents the use of conventional channel estimation
approaches \cite{Venugopal2017,SWOMP2018,Wu2019,Zhu2019}. The second stage exploits the geometric relationships between some of the
parameters of a sparse MIMO channel (angles and delays) and the position and orientation of the transmitter and the receiver to complete the localization.
Despite the success of compressive approaches, further work is needed to overcome obstacles to achieving higher localization accuracy along with relaxing key assumptions that make the algorithms more practical. To improve accuracy, more precise channel estimators are needed that do not suffer from high complexity encountered with larger dictionaries. To improve realizability, synchronization assumptions need to be relaxed. For example, prior work assumed that the absolute delay of the line of
sight (LoS) component can be directly obtained by the receiver
as part of the channel estimate, which requires an unrealistic
system where the transmitter and receiver are triggered at the
same time. To take advantage of larger arrays, more work is needed on hybrid architectures versus the fully digital architectures with high resolution converters, which is not feasible at mmWave frequencies due
to its high power consumption \cite{mmWavetutorial}. Finally, bandlimited models are needed to avoid an equivalent filtering effect based on a Dirac delta function, which leads to an artificial enhancement of the channel sparsity.

In this paper, we present  a practical approach to joint channel estimation and localization at mmWave that overcomes previous limitations. 
To drastically reduce complexity, we propose the use of a multidimensional orthogonal matching pursuit algorithm (MOMP)  \cite{MOMP} that operates with a dictionary in multiple dimensions instead of a large dictionary as conventional OMP \cite{Tropp2007} would do. To enable localization with a realistic transceiver, the position estimate is obtained as a function of the time difference of arrival and the angular parameters of the different multipath components, without relaying in an absolute time of arrival of the LoS path provided by the channel estimation algorithm. In addition, we also propose a novel approach to map channel parameters into position information by exploiting the special structure of reflections on walls, ceiling, and floor that appears in indoor channels. Numerical results in an indoor setting simulated by ray tracing show the effectiveness of the proposed strategy 
when the transmitter and receiver employ a practical hybrid MIMO architecture.

{\bf Notation:}
We use the following notation throughout the paper.
$x$, ${\bf x}$, ${\bf X}$ and $\mathcal{X}$ will be the styles for scalar, vector, matrix or tensor and set.
Regarding sub/supper-indices, $x$ and ${\rm x}$ are used to denote scalar and categorical values respectively.
$[{\bf x}]_n$ denotes the $n$-th entry of ${\bf x}$.
For a 2D matrix ${\bf X}$, $[{\bf X}]_{a, b}$, $[{\bf X}]_{a, :}$ and $[{\bf X}]_{:, b}$ are respectively, the element in the $a$-th row and $b$-th column, the $a$-th row and the $b$-th column, this notation is extended to the case of tensors with multi-indices acting like multiple indices, such as $[{\bf X}]_{{\bf a}, b} = [{\bf X}]_{a_1, a_2, b}$ for ${\bf a} = [a_1, a_2]$. The set  $\mathcal{U}^{n\times m}$ denotes the set of unit magnitude complex matrices of size $n$ by $m$.
We use the operator $\|{\bf x}\|$, $\|{\bf X}\|$ to denote the Euclidean and Frobenius norms of ${\bf x}$ and ${\bf X}$ respectively.
In this paper, indexing of tensors starts from 1 and we consider all vectors are column vectors. 

\section{System and signal model}
%

We consider a MIMO communication system with a hybrid architecture at both ends to enable operation at millimeter wave bands \cite{}. The number of antennas at the transmit array is $\Nt$, while  $\Nr$ denotes the number of antenna elements at the receive array.  The number of RF-chains is denoted as $\Mt$ and $\Mr$ for the transmitter and receiver respectively. We consider the transmission of $\Ns$ streams, with $\Ns \leq \Mt$ in general. During the link establishment phase, the transmitter sends a sequence of $M$ training symbol vectors to the receiver, so this can estimate the communication channel and the location of the transmitter.  The training symbols are transmitted by a sequence of training hybrid precoders designed to sound the channel, and received through a set of training hybrid combiners. For training purposes, we choose $\Ns=\Mt$, what leads to a square digital precoder.  The digital combiner is also chosen to be square during training, to avoid compressing the information captured by the receive RF chains.
With these assumptions in mind, the digital counterpart of the training precoders/combiners for the $m$-th training frame are $\FBBm \in\mathbb{C}^{\Mt\times \Mt}$ and $\WBBm\in\mathbb{C}^{\Mr\times \Mr}$. The analog training precoder and  the analog training combiner for the $m$-th training frame are denoted as $\FRFm \in\mathcal{U}^{\Nt \times \Mt}$  and $\WRFm \in \mathcal{U}^{\Nr\times \Mr}$. This way, the hybrid precoders/combiners are  $\Fm=\FRFm\FBBm\in\mathbb{C}^{\Nt\times \Mt}$ and
 $\Wm=\WRFm\WBBm\in\mathbb{C}^{\Nr\times \Mr}$. 
Regarding the construction of the training sequence itself, we consider the transmission of $\Mt$ streams during training, with a length $D$ zero padding and $Q$ symbols per stream and frame. $D$ is taken as the delay tap length of the channel. Under these assumptions and definitions, the training symbol matrix for the $m$-th frame is denoted as $\bS_m \in \mathbb{C}^{\Mt\times (Q+D)}$.

The frequency selective mmWave channel is modeled using a geometric channel model \cite{} with $L$ paths. The $d$-th delay tap of the channel, for $d=1,\ldots,D$, is represented as
\begin{equation}\label{eq:geometric_channel}
{\bf H}_{d} = \sum_{l = 1}^{L}\alpha_l\ar\atH p((d-1)T_{\rm s}+\tau_0-\tau_l),
\end{equation}
where $\alpha_l \in\mathbb{C}$, $\tau_l \in \mathbb{R}$, $\boldsymbol \theta_l$, and $\boldsymbol \phi_l$ are the complex gain, delay,  angle of arrival (AoA), and angle of departure (AoD) for the $l$-th path, $p(t)$ is a band limited pulse shaping function including filtering effects at the transmitter and receiver, $\tau_0$ is the delay between the beginning of
the transmission and the beginning of the reception (clock offset), and $\at$ and $\ar$
denote the array response vectors for the transmitter and the receiver.
The angular directions are represented as unitary vector directions, i.e. ${\boldsymbol \theta}_l, {\boldsymbol \phi}_l \in \{{\bf v}\in\mathbb{R}^3\text{ such that }\|{\bf v}\|=1\}$.

Because of the multiple array configurations exploited simultaneously with a hybrid architecture, the transmission of each training frame will generate a set of receive signals that we denote as the block matrix ${\bY}_m\in\mathbb{C}^{\Mr\times Q}$, comprised of $\Mr$ combinations of the $\Mt$ pilot signal streams. Considering a transmission power $P$, the expression of this received block matrix  is
\begin{equation}\label{eq:measurement_block}
[{\bY}_m]_{:, q} \! = \! \! \sqrt{P}\sum_{d = 1}^D\! \Wm^{\rm H}{\bf H}_d\Fm[{\bS_m}]_{:, q+D-d} \! + \! \Wm^{\rm H}[{\bN}_{m}]_{:, q},
\end{equation}
The noise block matrix ${\bf N}_{m}\in\mathbb{C}^{\Nr\times Q}$ has independent identically distributed entries following a distribution $\mathcal{NC}(0, \sigma^2)$, being $\sigma^2$ the noise power.
The final observation matrix $\bY$ is constructed by concatenating the received signals ${\bY}_m$ for the all the training frames, i.e. $\bY=[\bY_1, \bY_2, \ldots,\bY_M] \in \mathbb{C}^{\Mr \times QM}$. The problem to be solved consists of the estimation of the channel matrices ${\bf H}_d$ and the position of the transmitter assuming that the receiver position is fixed and known.

\section{MOMP-based mmWave channel estimation}\label{sec:channel_estimation}
Prior work  has developed a significant number of sparse recovery solutions to solve the problem of estimating the mmWave channel \cite{}. The proposed strategies incur, however,  in high computational complexity. This is mainly due to exploiting a formulation, for both the sensing matrix and the sparsifying dictionary, based on Kronecker products of large matrices. 
The Kronecker construction leads to an equivalent measurement matrix of significant dimension for common sizes of mmWave arrays, what makes matching pursuit solutions impractical in real-world 3D scenarios.   In this section, we derive a new formulation based on MOMP \cite{MOMP},  which leverages independent dictionaries in the angular and delay domains, without  building a global dictionary of larger size.

\subsection{Background in MOMP}
The multidimensional matching pursuit problem \cite{MOMP} consists of reconstructing a multidimensional sparse signal to best fit the available observations, assuming that it can be represented by projections on a given set of sparsifying dictionaries.
Let us first consider a whitened version of the observation matrix ${\bf Y}\in\mathbb{C}^{\Mr \times QM}$  as previously defined, which contains a set of $QM$ observations, each one of dimension $\Mr$. We denote the vectorized whitened measurements as ${\bf y} \in \mathbb{C}^{M \Mr Q}$.
To reconstruct a sparse tensor  from this observation using MOMP, we consider $N_{\rm D}$ dictionaries, with the $k$-th dictionary ${\bf \Psi}_k\in\mathbb{C}^{N_k^{\rm s}\times N_k^{\rm a}}$ consisting of $N_k^{\rm a}$ atoms in $\mathbb{C}^{N_k^{\rm s}}$. The coefficients of the sparse signal in the set of dictionaries are represented by ${\bf C}\in\mathbb{C}^{N_{1}^{\rm a}\times\ldots\times N_{N_{\rm D}}^{\rm a}}$. Our goal is to reconstruct the tensor containing these coefficients under the assumption that only a few elements are non-zero.
To cycle through the multiple indices of the dictionaries, we define the set of entry coordinate combinations $\mathcal{I} = \{{\bf i}=(i_1, \ldots, i_{N_{\rm D}})\in\mathbb{N}^{N_{\rm D}}\text{ s.t. }i_k \leq N_k^{\rm s}\quad\forall d \leq N_{\rm D}\}$, and the set of dictionary index combinations $\mathcal{J} = \{{\bf j}=(j_1, \ldots, j_{N_{\rm D}})\in\mathbb{N}^{N_{\rm D}}\text{ s.t. }j_k \leq N_k^{\rm a}, \forall d \leq N_{\rm D}\}$.
Finally, the sensing tensor ${\bf \Phi}\in\mathbb{C}^{M \Mr Q\times N_{1}^{\rm s}\times\ldots\times N_{N_{\rm D}}^{\rm s}}$ defines how the measurement of the sparse signal is performed. To recover the coefficients tensor we can use MOMP to solve:
\begin{equation}\label{eq:MOMP}
\min_{\bf C}\|{\bf y} - \sum_{{\bf i}\in\mathcal{I}}[{\bf \Phi}]_{:, {\bf i}} \left[\prod_{k=1}^{N_{\rm D}}{\bf \Psi}_{i_k, j_k^{1}}, \ldots, \prod_{k=1}^{N_{\rm D}}{\bf \Psi}_{i_k, j_k^{|\mathcal{J}|}}\right] {\bf C}_{\mathcal J}\|^2.
\end{equation}
Additionally, the support $\mathcal{C}\subset\mathcal{J}$ is defined as the set of indices such that any element of ${\bf C}_{\mathcal C}$ is non zero.

\subsection{MOMP-based formulation}
To formulate the mmWave channel estimation as a multidimensional orthogonal matching pursuit problem, we need to rewrite \eqref{eq:measurement_block} to match the structure of the problem in \eqref{eq:MOMP}. 
The next paragraphs derive the appropriate expressions for the dictionaries and the sensing matrix so that MOMP can be used to solve  \eqref{eq:MOMP} given the observation matrix ${\bY}$.

The channel tensor ${\bf H}\in\mathbb{C}^{\Nr\times \Mt\times D}$ is defined as $[{\bf H}]_{:, :, d}={\bf H}_{d}$.
To write the channel ${\bf H}$ as a sparse combination of products of dictionaries, it is natural to think of independent dictionaries for the angle of arrival, the angle of departure, and the delay, since these are the independent parameters of the paths that compose the channel. Moreover, two different dictionaries are used to represent the two independent dimensions of both the AoA and AoD.  With this in mind, we define five sparsifying dictionaries to represent the sparse mmWave channel.  

To obtain the compact expression of the dictionaries for the AoA, we consider that for a horizontal  uniform rectangular array of size $\Nr^{\rm x}\times \Nr^{\rm y}$, the array response vector $\ar\in\mathbb{C}^{\Nr^{\rm x}\Nr^{\rm y}}$ can be decomposed into two sub-components ${\bf a}_{\rm R}^{\rm x}(\theta^{\rm x})\in\mathbb{C}^{\Nr^{\rm x}}$ and ${\bf a}_{\rm R}^{\rm y}(\theta^{\rm y})\in\mathbb{C}^{\Nr^{\rm y}}$ such that $\ar = {\bf a}_{\rm R}^{\rm x}(\theta^{\rm x})\otimes{\bf a}_{\rm R}^{\rm y}(\theta^{\rm y})$.
This means that we can rewrite the entries of ${\bf a}_{\rm R}(\boldsymbol \theta_l)$ as $[{\bf a}_{\rm R}(\boldsymbol \theta_l)]_{\nr^{\rm x}\Nr^{\rm y}+\nr^{\rm y}} = [{\bf a}_{\rm R}^{\rm x}(\theta_l^{\rm x})]_{\nr^{\rm x}}[{\bf a}_{\rm R}^{\rm y}(\theta_l^{\rm y})]_{\nr^{\rm y}}$.
Analogously, to obtain the expression of the dictionaries for the AoD for a uniform rectangular array of size $\Nt^{\rm x}\times \Nt^{\rm y}$ at the transmitter,  we can find the decomposition $[{\bf a}_{\rm T}(\boldsymbol \phi_l)]_{\nt^{\rm x}\Nt^{\rm y}+\nt^{\rm y}} = [{\bf a}_{\rm T}^{\rm x}(\phi_l^{\rm x})]_{\nt^{\rm x}}[{\bf a}_{\rm T}^{\rm y}(\phi_l^{\rm y})]_{\nt^{\rm y}}$.
Finally, we will define the dictionary for the delay domain from the evaluation of the pulse shaping functions $[{\bf a}_{\rm D}(\tau)]_d = p((d-1)T_{\rm s}-\tau)$, $d=1,\ldots,D$.

Now, to obtain the final expressions for the dictionaries, we can consider the discrete domains for $\theta^{\rm x}$, $\theta^{\rm y}$, $\phi^{\rm x}$, $\phi^{\rm y}$ and $\tau-\tau_0$ with resolutions $N^{\rm a}_1$, $N^{\rm a}_2$, $N^{\rm a}_3$, $N^{\rm a}_4$ and $N^{\rm a}_5$, namely $\{\overline\theta_1^{\rm x}, \ldots, \overline\theta_{N^{\rm a}_1}^{\rm x}\}$, $\{\overline\theta_1^{\rm y}, \ldots, \overline\theta_{N^{\rm a}_2}^{\rm y}\}$, $\{\overline\phi_1^{\rm x}, \ldots, \overline\phi_{N^{\rm a}_3}^{\rm x}\}$, $\{\overline\phi_1^{\rm y}, \ldots, \overline\phi_{N^{\rm a}_4}^{\rm y}\}$ and $\{\overline\tau_1, \ldots, \overline\tau_{N^{\rm a}_5}\}$. This way, we define the dictionaries as
\begin{align}
&{\bf \Psi}_1 \! = \!  \left[{\bf a}_{\rm R}^{\rm x}(\overline\theta_1^{\rm x}), \ldots, {\bf a}_{\rm R}^{\rm x}(\overline\theta_{N_{1}^{\rm e}}^{\rm x})\right]\! \!, 
{\bf \Psi}_2 \! = \! \left[{\bf a}_{\rm R}^{\rm y}(\overline\theta_1^{\rm y}), \ldots, {\bf a}_{\rm R}^{\rm y}(\overline\theta_{N_{2}^{\rm e}}^{\rm y})\right]\! \!, \nonumber \\ 
&{\bf \Psi}_3 \! \! = \! \!  \left[{\bf a}_{\rm T}^{\rm x}(\overline\theta_1^{\rm x})^*, \ldots, {\bf a}_{\rm T}^{\rm x}(\overline\theta_{N_{3}^{\rm e}}^{\rm x})^*\! \right]\! \!, \! {\bf \Psi}_4 \! \! = \! \!  \left[{\bf a}_{\rm T}^{\rm y}(\overline\theta_1^{\rm y})^*, \ldots, {\bf a}_{\rm T}^{\rm y}(\overline\theta_{N_{4}^{\rm e}}^{\rm y})^*\! \right] \nonumber
\end{align} \vspace{-0.75cm}
\begin{align}
 {\bf \Psi}_5 =  \left[{\bf a}_{\rm D}(\overline\tau_1), \ldots, {\bf a}_{\rm D}(\overline\tau_{N_{5}^{\rm e}})\right].
\end{align}
To use these dictionaries, we define the sets of multi-indicies $\mathcal{I}=\{{\bf i}=[i_1, i_2, i_3, i_4, i_5]\in\mathbb{N}\text{ such that }i_1\leq \Nr^{\rm x}, i_2\leq \Nr^{\rm y}, i_3\leq \Nt^{\rm x}, i_4\leq \Nt^{\rm y}, i_5\leq D\}$, and $\mathcal{J}=\{{\bf j}=[j_1, j_2, j_3, j_4, j_5]\in\mathbb{N}\text{ such that }j_1\leq N_1^{\rm a}, j_2\leq N_2^{\rm a}, j_3\leq N_3^{\rm a}, j_4\leq N_4^{\rm a}, j_5\leq N_5^{\rm a}\}$ and $\Nr^{\rm x}\times \Nr^{\rm y}\times \Nt^{\rm x}\times \Nt^{\rm y}\times D$ can be expressed as $\otimes_{k=1}^5N_{k}^{\rm s}$.
Finally, ignoring quantization effects caused by the finite resolution of the dictionaries, we can define ${\bf C}_{\mathcal J}\in\mathbb{C}^{\otimes_{d=1}^{5}N_k^{\rm e} \times 1}$ as
the sparse tensor
\begin{equation}
{\bf C}_{\mathcal J}=\left\lbrace\begin{array}{cl}
\alpha_l, & \text{if } \begin{array}{ccc}
\theta_l^{\rm x}  = \overline\theta_{j_1}^{\rm x}\ {\rm or }\ \theta_l^{\rm y} =  \overline\theta_{j_2}^{\rm y} \ {\rm or} \ \\
\phi_l^{\rm x}  = \overline\phi_{j_3}^{\rm x} \ {\rm or }\  \phi_l^{\rm y} =  \overline\phi_{j_4}^{\rm y} \ {\rm or }\ \tau_l  = \overline\tau_{j_5}
\end{array}\\
0, & \text{otherwise.}
\end{array}\right.
\end{equation}
Under these definitions, the channel entries can now be expressed as
\begin{equation}\label{eq:channel_MOMP}
[{\bf H}]_{i_1\Nr^{\rm y} + i_2, i_3\Nt^{\rm y} + i_4, i_5} = \left[\prod_{k=1}^{N_{\rm D}}{\bf \Psi}_{i_k, j_k^{1}}, \ldots, \prod_{k=1}^{N_{\rm D}}{\bf \Psi}_{i_k, j_k^{|\mathcal{J}|}}\right]{\bf C}_{\mathcal J}.
\end{equation}

Regarding the construction of the observation, we assume that each training frame is independent. Next, the received signal is whitened to compensate for the correlation effect introduced by the analog combiner.   To this aim, we find a Cholesky decomposition of the noise correlation matrix ${\bf W}_m^{\rm H}{\bf W}_m$ i.e. ${\bf L}_m$ such that ${\bf L}_m{\bf L}_m^{\rm H} = {\bf W}_m^{\rm H}{\bf W}_m$ and multiply \eqref{eq:measurement_block} by ${\bf L}_m^{-1}$ to obtain 
\begin{multline}\label{eq:measurement_block_white}
[{\bf L}_m^{-1}{\bf Y}_m]_{:, q} = \sqrt{P}\sum_{d = 1}^D{\bf L}_m^{-1}{\bf W}_m^{\rm H}[{\bf H}]_{:, :, d}{\bf F}_m[{\bf S}_m]_{:, q+D-d}\\
+ {\bf L}_m^{-1}{\bf W}_m^{\rm H}[{\bf N}_m]_{:, q}.
\end{multline}
After the whitening stage, the expressions for the observation ${\bf y}$, and noise ${\bf n} \in\mathbb{C}^{M\MrQ}$ are
\begin{align}
[{\bf y}]_{mM_{\rm R}Q+m_{\rm R}Q+q} =& [{\bf L}_m^{-1}{\bf Y}_m]_{m_{\rm R}, q}\label{eq:y_definition}\\
[{\bf n}]_{mM_{\rm R}Q+m_{\rm R}Q+q} =& [{\bf L}_m^{-1}{\bf W}_m^{\rm H}{\bf N}_m]_{m_{\rm R}, q}.
\end{align}
To complete the MOMP formulation, we still need to define the sensing matrix ${\bf \Phi}\in\mathbb{C}^{M\MrQ\times \Nr^{\rm x}\times \Nr^{\rm y}\times \Nt^{\rm x}\times \Nt^{\rm y}\times D}$.
Note that we can also write the observations in \eqref{eq:y_definition} as $[{\bf y}]_{mM_{\rm R}Q+m_{\rm R}Q + q} =  [{\bf \Phi}]_{mM_{\rm R}Q+m_{\rm R}Q + q, {\bf i}} [{\bf H}]_{i_1\Nr^{\rm y} + i_2, i_3\Nt^{\rm y} + i_4, i_5}$. Mapping the terms in this observation to \eqref{eq:measurement_block_white}
\begin{multline}\label{eq:A}
[{\bf \Phi}]_{mM_{\rm R}Q+m_{\rm R}Q + q, {\bf i}} =\\
\sqrt{P}[{\bf L}_m^{-1}{\bf W}_m^{\rm H}]_{m_{\rm R}, i_1\Nr^{\rm y}+i_2}[{\bf F}_m{\bf S}_m]_{i_3\Nt^{\rm y}+i_4, q+D-i_5}.
\end{multline}
With these definitions, the multi-dimensional matching pursuit problem is completed, since we can re-write equation~\eqref{eq:measurement_block} like
\begin{equation}\label{eq:quasi-MOMP}
{\bf y} = \sum_{{\bf i}\in\mathcal{I}}[{\bf \Phi}]_{:, {\bf i}} \left[\prod_{k=1}^{N_{\rm D}}{\bf \Psi}_{i_k, j_k^{1}}, \ldots, \prod_{k=1}^{N_{\rm D}}{\bf \Psi}_{i_k, j_k^{|\mathcal{J}|}}\right] {\bf C}_{\mathcal J} + {\bf n}.
\end{equation}
Since ${\bf n}$ is white noise, the maximum likelihood estimator is the solution to \eqref{eq:MOMP}, and we can apply the MOMP algorithm to approximately and sparsely solve it.

The parameters for the estimated path corresponding to the index ${\bf j}\in\mathcal{C}$ can be directly extracted from the support and the sparse reconstructed matrix as $\overline\theta_{j_1}^{\rm x}$, $\overline\theta_{j_2}^{\rm y}$, $\overline\phi_{j_3}^{\rm x}$, $\overline\phi_{j_4}^{\rm y}$ and $\overline\tau_{j_5}$ for the angular information and relative time of arrival, and as ${\bf C}_{\mathcal J}$ for the complex gain.
To fully retrieve the angular information, the z components of $\boldsymbol \theta$ and $\boldsymbol  \phi$ can be computed using the fact that these are unitary vectors, and that the z component is positive (otherwise the path would be coming from the antenna array substrate which blocks the signal), i.e.  $\overline\theta^{\rm z}=\sqrt{(\overline\theta^{\rm x})^2+(\overline\theta^{\rm y})^2}$ and $\overline\phi^{\rm z}=\sqrt{(\overline\phi^{\rm x})^2+(\overline\phi^{\rm y})^2}$.

\section{Localization}
Once the channel parameters have been estimated, the geometric relationships between those parameters and the scatterers in the environment can be used to obtain 
an estimate of the transmitter position. 
These geometric relationships between the channel parameters and the position to be estimated depend on the type of path. Therefore, a method to classify the different estimated paths has to be proposed. To this aim, we will make use of two properties: a) specular reflections on the horizontal plane (floor/ceiling) do not change the $x$-$y$ components of a point, and b) vertical reflections (walls) do not change the $z$ component. With this in mind, we consider 4 possible types of paths: line of sight (LoS), wall reflection, floor/ceiling reflection, or any other path
that will be labeled as spurious and will not be exploited for localization.

The input to our proposed path classification algorithm are the spherical coordinates of the different angles, computed as  $\theta^{\rm az} = \arg(\theta^{\rm x} + j\theta^{\rm y})$, $\theta^{\rm el} = \arcsin(\theta^{\rm z})$, $\phi^{\rm az} = \arg(\phi^{\rm x} + j\phi^{\rm y})$ and $\phi^{\rm el} = \arcsin(\phi^{\rm z})$.
Wall reflections and LoS paths satisfy $\theta_l^{\rm el} + \phi_l^{\rm el} = 0$ while floor/ceiling reflections satisfy $\theta_l^{\rm el} = \phi_l^{\rm el}$.
Additionally, floor/ceiling reflections and LoS paths arrival and departure azimuth angles are opposite, i.e. they are separated by $180^\circ$.
By defining the threshold values $r_{\rm az}, r_{\rm el}$ we can obtain the conditions
\begin{align}
|\sin(\theta^{\rm el}-\phi_l^{\rm el})| < r_{\rm el}\label{eq:condition_el_v},\\
|\sin(\theta^{\rm el}+\phi_l^{\rm el})| < r_{\rm el}\label{eq:condition_el_h},\\
\cos(\theta^{\rm az}-\phi_l^{\rm az}) < r_{\rm az}-1.\label{eq:condition_az}
\end{align}
A LoS path satiifies conditions \eqref{eq:condition_el_v} and \eqref{eq:condition_az}, floor/ceiling reflections satisfy \eqref{eq:condition_el_h} and \eqref{eq:condition_az}, while wall reflections only satisfy  \eqref{eq:condition_el_v}. Any other path will be classified as spurious.

Let us define now the algorithm that provides the position estimate from the path parameters and  path classification. We consider the access point (receiver) to be in the coordinate origin, and ${\bf u}\in\mathbb{R}^3$ to be the device (transmitter) location.
To ease the formulation of the position estimation algorithm we introduce the projection matrices
${\boldsymbol \chi}_{\rm LOS} = {\bf I}_3$,
${\boldsymbol \chi}_{\rm h} = \left[\begin{array}{cc}
1 & 0\\
0 & 1\\
0 & 0
\end{array}\right]$, and
${\boldsymbol \chi}_{\rm v} = \left[\begin{array}{c}
0\\
0\\
1
\end{array}\right]$. With these definitions, the geometric relationships created by a LoS path, a floor/ceiling reflection or a wall reflection can be expressed as
\begin{align}
{\boldsymbol \chi}_{\rm LOS}{\bf u} = \! {\boldsymbol \chi}_{\rm LOS}{\boldsymbol \theta}_l\tau_l,\ 
{\boldsymbol \chi}_{\rm h}{\bf u} = \! {\boldsymbol \chi}_{\rm h}{\boldsymbol \theta}_l\tau_l,\ 
 {\boldsymbol \chi}_{\rm v}{\bf u} = \! {\boldsymbol \chi}_{\rm v}{\boldsymbol \theta}_l\tau_l,
\end{align}
respectively. In other words, if we define ${\boldsymbol \chi}_l$, as the projection matrix corresponding to the $l$-th path type, we can rewrite the geometric relationships for that path as
\begin{equation}
{\boldsymbol \chi}_l({\bf u} - {\boldsymbol \theta}_l\tau_l) = 0.
\end{equation}
Note, however, that the channel estimation algorithm only provides the estimation of the relative delays $\Delta\tau_l = \tau_l-\tau_0$, with $\tau_0$ unknown,  instead of $\tau_l$. 
Therefore, the equations for the geometric relationships need to be established in terms of the relative delays, i.e.
\begin{equation}
{\boldsymbol \chi}_l({\bf u} - {\boldsymbol \theta}_l(\Delta\tau_l + \tau_0)) = 0.
\label{eq:geometry}
\end{equation}
Now, we define the matrix ${\bf U}_{{\boldsymbol \theta}_l} = [{\bf I}, -{\boldsymbol \theta}_l]$ and  the variable ${\bf z} = [{\bf u}^{\rm T}, \tau_0]^{\rm T}$, which contains all the unknown variables, to rewrite \eqref{eq:geometry}  as
\begin{equation}
{\boldsymbol \chi}_{\rm p}({\bf U}_{{\boldsymbol \theta}_l}{\bf z} - {\boldsymbol \theta}_l\Delta\tau_l) = 0.
\end{equation}
This equation can be alternatively solved by 
\begin{equation}\label{eq:localization_path}
\min_{{\bf z}}{\bf z}^{\rm T}{\bf U}_{{\boldsymbol \theta}_l}^{\rm T}{\boldsymbol \chi}_l^{\rm T}{\boldsymbol \chi}_l{\bf U}_{{\boldsymbol \theta}_l}{\bf z} - 2\tau_l\phi_l^{\rm T}{\boldsymbol \chi}_l^{\rm T}{\boldsymbol \chi}_l{\bf U}_{{\boldsymbol \theta}_l}{\bf z} + \tau_l^2\phi_l^{\rm T}{\boldsymbol \chi}_l^{\rm T}{\boldsymbol \chi}_l\phi_l.
\end{equation}
We define now a weighted version of the previous equation to account for the different impact of the errors associated to different types of paths. Thus, we consider a weight value for each path $w_l$ that we define as its estimated gain $w_l = |\alpha_l|^2$.
With this in mind we can weight average \eqref{eq:localization_path} for all the different paths as
\begin{equation}\label{eq:localization_quad}
\min_{{\bf z}}{\bf z}^{\rm T}{\bf A}{\bf z} - 2{\bf b}{\bf z} + c,
\end{equation}
for ${\bf A} = \sum_{l=1}^Lw_l{\bf U}_{{\boldsymbol \theta}_l}^{\rm T}{\boldsymbol \chi}_l^{\rm T}{\boldsymbol \chi}_l{\bf U}_{{\boldsymbol \theta}_l}$, ${\bf b} = \sum_{l=1}^Lw_l\tau_l\phi_l^{\rm T}{\boldsymbol \chi}_l^{\rm T}{\boldsymbol \chi}_l{\bf U}_{{\boldsymbol \theta}_l}$ and $c = \sum_{l=1}^Lw_l\tau_l^2\phi_l^{\rm T}{\boldsymbol \chi}_l^{\rm T}{\boldsymbol \chi}_l\phi_l$.
The widely known solution to \eqref{eq:localization_quad} for a symmetric ${\bf A}$ matrix is $\hat{\bf z} = {\bf A}^{-1}{\bf b}$.
The location estimation $\hat{\bf u}$ can then be extracted from the first 3 entries of $\hat{\bf z}$, while the last entry provides the unkown offset between transmitter and receiver.

\section{Simulation}
To generate the evaluation data set with channels and user positions we simulate by ray tracing (using Wireless InSite sofware)  a home office with two rooms. 
There is an access point in each one of the rooms. We generate $218$ user locations following a path connecting both rooms.
The user antenna is place horizontally, while the access points are mounted on the walls vertically, facing the interior of their respective rooms.
The ray tracing software generates the optical geometrical channel between the user positions and the access points.
We associate each user location to the single access point with the highest received signal strength.
For each user location, we input the parameters corresponding to the channel with its associated access point in a measurement model to generate the received signal.
The  transmit power is set to $P = 20{\rm dBm}$, and the noise level is the thermal noise corresponding to a room at $15^\circ {\rm C}$ when using a $1{\rm GHz}$ bandwidth, i.e. $\sigma^2 = -84{\rm dBm}$. 
The pulse shaping filter is defined as a sinc function.
We set the delay tap length to $D = 64$, and the training signal to be the first $6$ rows of a $64$ element Hadamard matrix padded with $64$ zeros to the left and $32$ zeros to the right, i.e. ${\bf S} = [0, \ldots, 0, {\bf Had}_{64}, 0, \ldots, 0]$.
The training precoders/combiners are taken as columns extracted from the Kronecker product of the DFT matrices with sizes $N_{\rm T}^{\rm x}$ and $N_{\rm T}^{\rm y}$/$N_{\rm R}^{\rm x}$ and $N_{\rm R}^{\rm y}$. 


To simplify the sparse reconstruction problem, we define the ratio between the number of atoms and atoms size $K_{\rm res}=N_{k}^{\rm a}/N_{k}^{\rm s}$, setting it to the same constant value independent of $k$.
By how the dictionaries are defined we have $N_1^{\rm s} = N_{\rm R}^{\rm x}$, $N_2^{\rm s} = N_{\rm R}^{\rm y}$, $N_3^{\rm s} = N_{\rm T}^{\rm x}$, $N_4^{\rm s} = N_{\rm R}^{\rm y}$, and $N_5^{\rm s}=D$. 

The first metric used to evaluate the accuracy of the channel estimation algorithm is the error in the estimation of the AoA of the main path.
We can see its evolution over $K_{\rm res}$ for both algorithms, OMP and MOMP, in Table~\ref{tab:KvAngle}, when considering a $3\times 3$ antenna array at the transmitter and a $6\times6$ antenna at the receiver. The number of RF chains is set to 3 for the device and to 6 for the access point.
\begin{table}[t!]
\begin{center}
\begin{tabular}{|c|c|c|c|c|c|c|}\hline
Method & \multicolumn{3}{c|}{OMP} & \multicolumn{3}{c|}{MOMP}\\\hline
$K_{\rm res}$ & $1$ & $1.3$ & $1.6$ & $16$ & $128$ & $1024$\\\hline
$\angle(\hat{\boldsymbol \theta}_1,{\boldsymbol \theta}_1) [^\circ]$ & $12$ & $10$ & $8$ & $1.1$ & $0.9$ & $0.9$\\\hline
$\angle(\hat{\boldsymbol \phi}_1, {\boldsymbol \phi}_1) [^\circ]$ & $44$ & $30$ & $25$ & $2.1$ & $1.8$ & $1.7$\\\hline
Run-time $[{\rm s}]$ & $1.3$ & $2.2$ & $8.2$ & $3.2$ & $3.3$ & $3.5$\\\hline
\end{tabular}
\end{center}
\caption{Median main path angular error and run time as a function of $K_{\rm res}$.}
\label{tab:KvAngle}
\end{table}
It is straight forwards to see how MOMP outperforms OMP estimations by an order of magnitude because of its ability to increase the resolution of the dictionaries without incurring in exorbitant computational times or impossible memory requirements. 

\begin{figure}
\begin{center}
\includegraphics[width=0.8\linewidth]{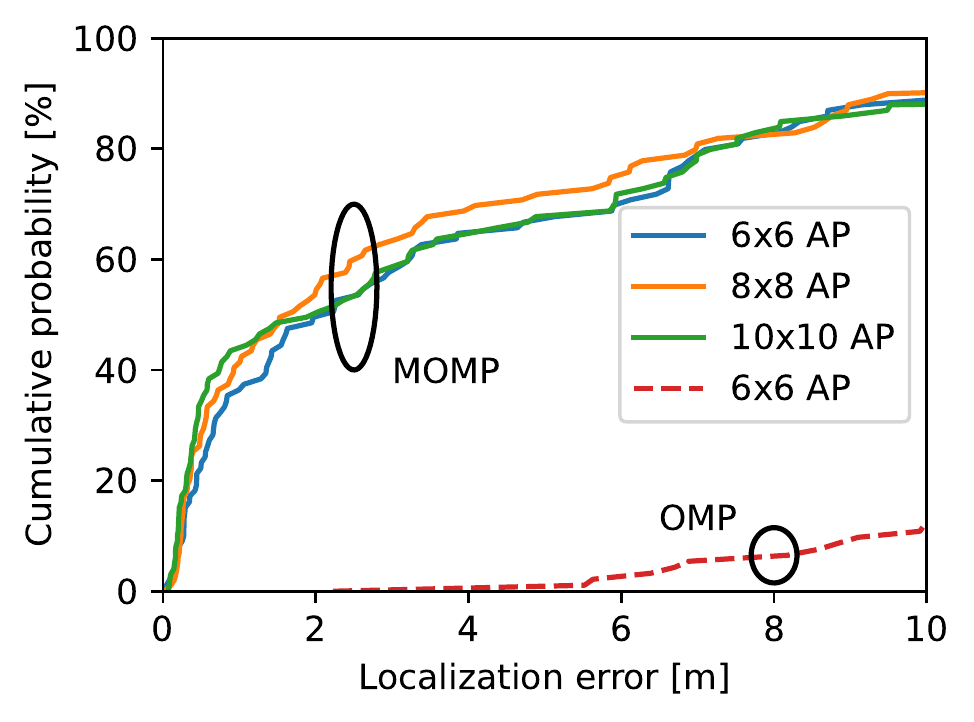}
\end{center}
\caption{Localization error as a function of the receive array size, number of RF chains and sparse recovery strategy. $K_{\rm res}$ has been set to 128 for MOMP and to 1.6 for OMP.}
\label{fig:Loc_Error}
\end{figure}

We analyze next the localization error $\|{\bf u}-\hat{\bf u}\|$ as a function of the number of antennas and complexity. The value used for the weight is each path's estimated power $w_l = |\bar\alpha_l|^2$ and the threshold values for the path classification are $r_{\rm az} = r_{\rm el} = 0.12$. We fix the antenna size for the device to $3\times3$, while three different sizes are considered for the access points: $6\times 6$, $8\times 8$, and $10\times 10$. The number of receive RF chains also varies accordingly as 6, 8 and 10.
The localization results can be visualized in Fig.~\ref{fig:Loc_Error}. Even with these  small antennas sizes, the access point is able to localize most of the devices keeping the error under reasonable values when an unknown clock offset is considered in the simulation. Note that even channels with very bad SNR conditions, where the link cannot be established have been considered in these averaged results. 
These results clearly show the benefits of applying MOMP to the problem of joint channel estimation and localization at mmWave. Higher estimation accuracies are expected with larger antenna arrays at the transmitter, which is constrained to be only $3\times3$ in this simulation scenario so OMP can run.

\section{Conclusions}
We developed a joint compressive channel estimation and localization strategy for a realistic mmWave MIMO communication systems.
The proposed approach is based on a multidimensional matching pursuit sparse recovery algorithm, which enables operation with high resolution dictionaries, for both the time and angular domains, at a reduced complexity. Using mmWave channels generated by ray tracing we obtained the channel estimation error for the angular parameters and the localization errors for a random deployment of users, as a function of the complexity and the system parameters. We showed how the MOMP-based 
 approach is able to provide reasonable estimation accuracies in scenarios where conventional OMP is not feasible.

\bibliographystyle{IEEEtran}  
\bibliography{refs}     


\end{document}